\documentstyle[11pt,epsfig]{article}
\title{\bf Unification of Higgs  and Maxwell fields in Brane-Kaluza-Klein
gravity}
\author{S. Jalalzadeh$^1$\thanks{email: s-jalalzadeh@cc.sbu.ac.ir}
and H. R. Sepangi$^{1,2}$\thanks{email: hr-sepangi@cc.sbu.ac.ir}
\\ $^1${\small Department of Physics, Shahid Beheshti University, Evin,
Tehran 19839, Iran}\\$^2${\small
Institute for Studies in Theoretical Physics and Mathematics, P.O.
Box 19395-5746, Tehran, Iran }}
\begin{document}
\maketitle 

\begin{abstract}The unification of  Higgs and
electromagnetic fields in the context of higher dimensional
gravity is studied. We show that these fields arise from an extra
large dimension together with a compact small dimension. The
question of the localization of the gauge fields  and their
relation to junction conditions is also addressed.
\vspace{0.5cm}
\end{abstract}\pagebreak
\section{Introduction}
One of the important, but unanswered questions of the physics of
elementary particles relates to the origin of the spectrum of the
known particles.  In particular, the observed hierarchy of
particle masses still suffers from the absence of a coherent and
fundamental understanding. Promising ideas in this regard for the
past decades have been mainly based on symmetry principles. The
idea of spontaneous symmetry breaking embodied in the appearance
of the Higgs field is the fruit of such attempts from which the
contemporary physics has been greatly benefited. The Higgs field
therefore, has been playing an undeniably important role in the
realm of particle physics and cosmology ever since the idea came
into fruition in the 1960's. Because of its success, people have
been trying to devise mechanisms in order to give a plausible
explanation for its origin.

Generally  speaking, in quantum field theory, the potential
representing the Higgs particle enters the theory in a rather
spurious way  and its origin is therefore not well explained.
However, in recent years, theories of gravity with extra
dimensions have become an important tool, both in particle physics
and  cosmology and been used in order to gain more understanding
in the nature and origin of fundamental particles and forces. For
example, extra compact dimensions have been used to address the
hierarchy problem \cite{arkani}; the huge disparity in size
between various fundamental constants in nature. In another
approach, large extra dimensions \cite{randal} have been used to
reduce the unification scale from Plankian to that of Tev.
Interestingly, these theories have also been used in the past few
years to account for the origin of the Higgs field. Indeed, the
Higgs field is studied by considering the extra dimensional
components of higher dimensional gauge fields in five dimensions
in \cite{burdman}, or the Higgs hierarchy problem is dealt with in
\cite{benakli} by considering the transverse (submillimeter) and
longitudinal (Tev) extra dimensions.

In this paper, we study the Higgs field in a theory with two extra
dimensions, one being compact and the other large. For the compact
dimension the theory is of the Kaluza-Klien type and for the large
extra dimension it follows the footsteps of the brane theory. The
compact extra dimension is characterized by the gauge fields
$A_\mu$ and  the scalar field $\phi$. A perturbation of the
coordinates of the bulk space in terms of the coordinates of the
brane and the normal direction together with Gauss-Codazzi
equations and the imposition of the Israel junction conditions
culminates in an effective action for the theory in which the
potential turns out to become that of the Higgs having a mass
characterized by the constants of the theory. It is interesting to
note that we have made no {\it apriori} assumptions about the
fields and their properties. The characteristic of these fields
have been deduced from purely geometrical considerations.
\section{Geometrical setup}
Consider the background manifold $\overline{V}_{5}$ isometrically
embedded in $ V_{6}$ by the map ${ \cal Y}:
\overline{V}_{5}\rightarrow  V_{6}$ such that \footnote{Capital
Latin indices run from 0 to 5, small Latin indices run from 0 to 3
and Greek indices run from 0 to 4. Quantities representing the
bulk are distinguished from that of the brane by calligraphic
letters.}
\begin{eqnarray}
{\cal G} _{AB} {\cal Y}^{A}_{,\mu } {\cal Y}^{B}_{,\nu}=
\bar{g}_{\mu \nu}  , \hspace{.5 cm} {\cal G}_{AB}{\cal
Y}^{A}_{,\mu}{\cal N}^{B} = 0  ,\hspace{.5 cm}  {\cal G}_{AB}{\cal
N}^{A}{\cal N}^{B} = \epsilon,\hspace{.5 cm} \epsilon = \pm 1
\nonumber
\end{eqnarray}
where $ {\cal G}_{AB} $  $ ( \overline{g}_{\mu\nu} ) $ is the
metric of the bulk (brane) space $V_{6}$  $( \overline{V}_{5} )$,
$ \{ {\cal Y}^{A} \} $ and  $\{ x^{\mu}\} $ are the bases
expanding the bulk (brane) and  ${\cal N}^{A} $ is a normal unit
vector orthogonal to the brane.  In this paper we adapt the
Kaluza-Klein metric as the  background metric and write it as
\begin{eqnarray}
\bar{g}_{\mu\nu} = \left(\!\!\!\begin{array}{cc}
  \bar{g}^{(0)}_{ij}( x^{l} ) + \phi^{ 2 } A_{ i }A_{ j } & \phi ^{2} A_{ j } \\
  \phi^{ 2 }A_{ i } & \phi^{2}
\end{array}\!\!\!\right) \label{eq1}
\end{eqnarray}
where $\bar{g}^{(0)}_{ij}$ is the metric of the $4D$  spacetime ,
$A_{i}$ is the Maxwell gauge field and $\phi$ is the scale factor
of the small compact dimension. The integrability condition for
the background geometry are given by the Gauss-Codazzi equations
which can be written as \cite{eis}
\begin{eqnarray}
\bar{R}_{\mu\nu\alpha\beta} = 2\bar{K}_{ \mu [ \beta }\bar{ K
}_{\alpha ] \nu } + {\cal R}_{ABCD}{\cal Y}^{A}_{,\mu}{\cal
Y}^{B}_{,\nu}{\cal Y}^{C}_{,\alpha}{\cal Y}^{D}_{,\beta}
\label{eq2}
\end{eqnarray}
and
\begin{eqnarray}
\bar{K}_{\alpha[ \beta ; \gamma ]} = {\cal R}_{ABCD}{\cal
Y}^{A}_{,\alpha}{\cal N}^{B}{\cal Y}^{C}_{,\beta}{\cal
Y}^{D}_{,\gamma} \label{eq3}
\end{eqnarray}
where ${\cal R}_{ABCD}$  $(\bar{R}_{\mu\nu\alpha\beta})$ is the
Riemann curvature of bulk (brane) manifold and
\begin{eqnarray}
\bar{K}_{\alpha\beta} = {\cal G}_{AB} \left( {\cal
Y}^{A}_{,\alpha\beta} + \Gamma ^{A}_{CD}  {\cal Y}^{C}_{,\alpha}
{\cal Y}^{D}_{,\beta} \right) {\cal N}^{B} \nonumber
\end{eqnarray}
is the extrinsic curvature.

 At this stage we impose the cylinder condition, namely $
 \bar{g}_{\mu\nu,4} = 0$. This means that the fields $
 \bar{g}^{(0)}_{ij}( x )$, $ A_{\mu}( x ) $ and $ \phi( x ) $
 are independent of the fifth coordinate $ x^{4} $. In a
 sufficiently  small neighborhood of $ \overline{V}_{5} $,
 a local coordinate of the bulk space can be written as
 \cite{maia}
 \begin{equation}
 {\cal Z}^{A} = {\cal Y}^{A} + s {\cal N}^{A} \label{eq4}
 \end{equation}
where $s$ is a small parameter along ${\cal N}^{A}$ that
parameterizes the large non compact extra dimension. In the
adapted coordinate (\ref{eq4}) the components of the bulk space
metric can be rewritten as
\begin{eqnarray}
\gamma_{\alpha\beta} = {\cal G}_{AB}{\cal Z}^{A}_{,\alpha}{\cal
Z}^{B}_{,\beta}  \hspace{.3 cm} \gamma_{\alpha 5} = {\cal
G}_{AB}{\cal Z}^{A}_{,\alpha}{\cal N}^{B} \hspace{.3 cm}
\gamma_{55} = {\cal G}_{AB}{\cal N}^{A}{\cal N}^{B}. \nonumber
\end{eqnarray}
If we set ${\cal N}^{A} = \delta^{A}_{5}$, the metric of the bulk
space can then be written in the form
 \begin{eqnarray}
{\cal G}_{AB} = \left(\!\!\!\begin{array}{cc}
 g_{\alpha\beta}& 0 \\
 0 &\epsilon
\end{array}\!\!\!\right) \label{eq5}
\end{eqnarray}
where in terms of the original brane metric and the corresponding
extrinsic curvature, $g_{\alpha\beta}$ are given by
\begin{eqnarray}
g_{\alpha\beta} = \bar{g}_{\alpha\beta} - 2 s
\bar{K}_{\alpha\beta} +
s^{2}\bar{g}^{\mu\nu}\bar{K}_{\alpha\mu}\bar{K}_{\beta\nu}.
\label{eq6}
\end{eqnarray}
It should be clear that each $s =\mbox{const}.$ represents a
(perturbed) brane that is isometrically embedded in a bulk for
which the integrability conditions (\ref{eq2})  read
\begin{eqnarray}
R_{\mu\nu\alpha\beta} = 2K_{ \mu [ \beta }K_{\alpha ] \nu } +
{\cal R}_{ABCD}{\cal Z}^{A}_{,\mu}{\cal Z}^{B}_{,\nu}{\cal
Z}^{C}_{,\alpha}{\cal Z}^{D}_{,\beta} \label{eq7}\\
\nonumber \\
K_{\alpha[ \beta ; \gamma ]} = {\cal R}_{ABCD}{\cal
Z}^{A}_{,\alpha}{\cal N}^{B}{\cal Z}^{C}_{,\beta}{\cal
Z}^{D}_{,\gamma}\label{eq8}
\end{eqnarray}
with the extrinsic curvature now given by
\begin{eqnarray}
K_{\alpha\beta} = -\frac{1}{2}\frac{\partial
g_{\alpha\beta}}{\partial s} = \bar{K}_{\alpha\beta} - s
\bar{g}^{\mu\nu}\bar{K}_{\mu\alpha}\bar{K}_{\nu\beta}. \label{eq9}
\end{eqnarray}
At this point, it is desirable to calculate the Ricci scalar and
Ricci tensor corresponding to our perturbed brane as this would
enable us to define an effective action describing the theory. To
this end, contraction of equation (\ref{eq7}) leads to
\begin{equation}
R_{\alpha\beta} = 2\epsilon g^{\mu\nu}K_{\mu[\beta}K_{\nu]\alpha}
+ g^{\mu\nu}{\cal R}_{ABCD}{\cal Z}^{A}_{,\mu}{\cal
Z}^{C}_{,\nu}{\cal Z}^{B}_{,\alpha}{\cal Z}^{D}_{,\beta}.
\label{eq10}
\end{equation}
Using the identity
\begin{eqnarray}
g^{\mu\nu}{\cal Z}^{A}_{,\mu}{\cal Z}^{B}_{,\nu} = {\cal G} ^{AB}
- \epsilon {\cal N}^{A}{\cal N}^{B} \nonumber
\end{eqnarray}
the Ricci tensor and Ricci scalar of the perturbed brane are
given by
\begin{eqnarray}
R_{\alpha\beta} = {\cal R}_{AB}{\cal Z}^{A}_{,\alpha}{\cal
Z}^{B}_{,\beta} - 2 \epsilon g^{\mu\nu} K_{\mu[\nu}
K_{\alpha]\beta} -\epsilon {\cal R}_{ABCD}{\cal
Z}^{B}_{,\alpha}{\cal Z}^{D}_{,\beta} {\cal N}^{A}{\cal N}^{C}
\label{eq11}
\end{eqnarray}
and
\begin{eqnarray}
R = {\cal R} + \epsilon K_{\mu\nu}K^{\mu\nu} -\epsilon K^{2} - 2
\epsilon {\cal R}_{AB}{\cal N}^{A}{\cal N}^{B} - {\cal
R}_{ABCD}{\cal N}^{A}{\cal N}^{B}{\cal N}^{C}{\cal N}^{D}
\label{eq12}
\end{eqnarray}
where $ K = g^{\mu\nu} K_{\mu\nu} $is the mean curvature. With a
straightforward calculation in the Gaussian frame of the embedding
represented by (\ref{eq5}) we are led to the following equations
\begin{eqnarray}
{\cal R}_{ABCD}{\cal Z}^{B}_{,\alpha}{\cal Z}^{D}_{,\beta}{\cal
N}^{A}{\cal N}^{C} = - \epsilon K_{\alpha\mu}K^{\mu}_{\beta} -
\epsilon K_{\alpha\beta,s} \label{eq13} \\
\nonumber \\
{\cal R}_{AB}{\cal N}^{A}{\cal N}^{B} = \epsilon
K_{\mu\nu}K^{\mu\nu} -\epsilon K_{,s} \label{eq14} \\
\nonumber \\
{\cal R}_{ABCD}{\cal N}^{A}{\cal N}^{B}{\cal N}^{C}{\cal N}^{D} =
0 \label{eq15}
\end{eqnarray}
so that the Ricci tensor and Ricci scalar of the perturbed brane
take the following forms
\begin{eqnarray}
R_{\alpha\beta} = {\cal R}_{AB}{\cal Z}^{A}_{,\alpha}{\cal
Z}^{B}_{,\beta} +\epsilon \left( 2 K_{\alpha\mu} K^{\mu}_{\beta} -
K K_{\alpha\beta} \right)  + \epsilon K_{\alpha\beta , s}
\label{eq16}
\end{eqnarray}
and
\begin{eqnarray}
R = {\cal R} - \epsilon \left( K^{2} + K_{\mu\nu} K^{\mu\nu}
\right) - 2\epsilon K_{, s}. \label{eq17}
\end{eqnarray}
\section{The model}

Equation (\ref{eq17}) suggests that the effective action for the
brane-world geometry can now be written as
\begin{eqnarray}
{\cal S}_{eff} &=&\frac{1}{2K^{2}_{(b)}} \int \sqrt{-g}R d^5 x
=\frac{1}{2K^{2}_{(b)}} \int \sqrt{-g}\left[{\cal R}-\epsilon
\left( K^{2} + K_{\mu\nu}K^{\mu\nu}\right)  - 2\epsilon K_{, s}
\right] d^5x \label{eq18}
\end{eqnarray}
where $K_{(b)}$ is the gravitational constant of the brane world.
In the spirit of the variational principle and following
\cite{maia}, we add an independent source as the Lagrangian of the
confined matter $( {\cal L}_{m} ) $ together with the
corresponding tension of the $5D$ brane $ ( \sigma ) $ to our
geometrical Lagrangian in the form ${\cal L}_m-2\sigma$.
 By using the variational principle or directly from equation (\ref{eq16})
and (\ref{eq17}), the field equations become
\begin{eqnarray}
G_{\mu\nu} &=& G_{AB} {\cal Z}^{A}_{\mu}{\cal Z}^{B}_{\nu} -
\epsilon K K_{\mu\nu} + 2 \epsilon K_{\mu\alpha} K^{\alpha}_{\nu}
 +
\epsilon K_{\mu\nu ,s}+\frac{\epsilon}{2} K^{2} g_{\mu\nu} \nonumber\\
&+& \frac{\epsilon}{2} K_{\alpha\beta} K^{\alpha\beta} g_{\mu\nu}-
\epsilon K_{, s} g_{\mu\nu} + K^{2}_{(b)}S_{\mu\nu} \label{eq19}
\end{eqnarray}
where $ S_{\mu\nu} $ is the matter energy-momentum tensor
associated with the brane-world. This tensor consists of three
parts
\begin{eqnarray}
S_{ij} = \tau_{ij} - \sigma g^{(4)}_{ij} + A_{i}A_{j}\phi^{2}
\left(
{\cal L}_{m} - \sigma \right) \label{eq20} \\
\nonumber\\
S_{i4} = \phi^{2}A_{i} \left( {\cal L}_{m} - \sigma \right) \label{eq21} \\
\nonumber \\
S_{44} = \phi^{2} \left({\cal L}_{m} - \sigma \right) \label{eq22}
\end{eqnarray}
with $$ \tau_{ij}=-\frac{2}{\sqrt{-g}}\frac{\delta({\cal
L}\sqrt{-g})}{\delta g^{ij}}.$$ In order to obtain the effective
field equations on the brane, we have to replace the terms
$\epsilon K_{, s}\, g_{\mu\nu} $ and $ \epsilon K_{\mu\nu, s} $ in
equation (\ref{eq19}) with the $5D$ variables on the perturbed
brane. Decomposing the $6D$ Riemann tensor as
\begin{eqnarray}
{\cal R}_{ABCD} = {\cal C}_{ABCD} +\frac{1}{2}\left( {\cal
G}_{A[C}{\cal R}_{D]B} - {\cal G}_{B[C} {\cal R}_{D]A} \right)
-\frac{1}{10} {\cal G}_{A[C} {\cal G}_{D]B}{\cal R} \label{eq23}
\end{eqnarray}
where $ {\cal C}_{ABCD} $ is the $6D$ Weyl curvature, we find
\begin{eqnarray}
K_{\mu\nu, s} -\frac{1}{5} g_{\mu\nu} K_{, s} &=&-
\frac{4}{3}{\cal E}_{\mu\nu} - \frac{2}{3} K_{\mu\alpha}
K^{\alpha}_{\nu} - \frac{1}{5} K_{\alpha\beta} K^{\alpha\beta}
g_{\mu\nu} - \frac{\epsilon}{3} R_{\mu\nu} + \frac{\epsilon}{15} R
g_{\mu\nu} \nonumber\\ &-& \frac{1}{3} K K_{\mu\nu} + \frac{1}{15}
K^{2} g_{\mu\nu} \label{eq24}
\end{eqnarray}
where $ {\cal E}_{\mu\nu} = {\cal C}_{ABCD}{\cal N}^{A}{\cal
N}^{C}{\cal Z}^{B}_{, \mu}{\cal Z}^{D}_{, \nu} $ is the electric
part of the Weyl tensor. Since $ {\cal E}_{\mu\nu} $ is traceless,
we can not fix $ K_{\mu\nu, s} $ by equation (\ref{eq24}). To
address this problem, we assume that the field equations on the
bulk represent either a de-Sitter or anti de-Sitter space with a
cosmological constant denoted by $\Lambda^{(B)}$, namely
\begin{eqnarray}
G_{AB} = K^{2}_{( B )} \left(-\Lambda^{( B )}{\cal G}_{AB} + T^{(
b )}_{AB} \right) \label{eq25}
\end{eqnarray}
where
\begin{eqnarray}
T^{(b)}_{AB} = \delta^{\mu}_{A}\delta^{\nu}_{B} S_{\mu\nu} \delta(
s ).
\end{eqnarray}
where $K_{(B)}$ is the gravitational constant of the bulk space.
Now, taking the trace of equation (\ref{eq25}), we find
\begin{eqnarray}
{\cal R} = 3\Lambda^{(B)} K^{2}_{(B)}. \label{eq26}
\end{eqnarray}
Using equations (\ref{eq17}), (\ref{eq26}) and (\ref{eq24}) the
field equations (\ref{eq19}) can be  rewritten as
\begin{eqnarray}
G_{\mu\nu} = K^{2}_{(b)} S_{\mu\nu} - \frac{3}{4} \Lambda^{(B)}
K^{2}_{(B)} g_{\mu\nu} - \epsilon K K_{\mu\nu} + \epsilon
K_{\mu\alpha} K^{\alpha}_{\nu} - \frac{1}{2} \epsilon \left(
K_{\mu\nu} K^{\mu\nu} - K^{2} \right) g_{\mu\nu}. \label{eq27}
\end{eqnarray}
Equation (\ref{eq27}) represents the field equations on the brane
written in terms of the extrinsic curvature. However, the form of
the field equations in our $4D$ world would also be of much
interest and would tell us about the character  of the fields
$A_i$ and $\phi$ in terms of which the metric (\ref{eq1}) was
introduced. To achieve this, we invoke Israel junction conditions.
These junction conditions are obtained by assuming $Z_2$ symmetry
and substituting  equation (\ref{eq25}) into (\ref{eq19}) and
integrating the resulting equation in the normal direction to the
brane, with the result
\begin{eqnarray}
K_{\mu\nu} = -\frac{1}{2}\epsilon K^{2}_{(B)} \left[ S_{\mu\nu}
-\frac{1}{4} S g_{\mu\nu} \right] \label{eq227}
\end{eqnarray}
where S is the trace of $ S_{\mu\nu} $.

To calculate the $4D$ field equations we first need to calculate
the $4D$ part of the Ricci tensor and Ricci scalar of the $5D$
spacetime in the $\{ i , 4 \}$ coordinate using metric
(\ref{eq1}). The results are the following equations \cite{finn}
\begin{eqnarray}
R_{44} &=& \frac{1}{4}\phi^{4} F_{\mu\nu}F^{\mu\nu} - \phi
\nabla^{2}\phi \label{eq28} \\
R_{i4} &=& \frac{1}{2} \phi^{2}\nabla^{i} F_{ij} + \frac{3}{2}
\phi
F_{ij} \nabla^{j} \phi + A_{i} R_{44} \label{eq29}\\
 R_{ij} &=& R^{(4)}_{ij} -\frac{1}{2} \phi^{2}F^{m}_{i}F_{mj}
- \frac{1}{\phi}\nabla_{i}\nabla_{j}\phi + A_{i}A_{j}R_{44} +
\frac{1}{2}\phi A_{i} \left( \phi \nabla^{l}F_{jl} +
3F_{jl}\nabla^{l}\phi \right)\nonumber \\ &+&\frac{1}{2}\phi A_{j}
\left(
\phi \nabla^{l}F_{il} + 3 F_{il}\nabla^{l}\phi \right) \label{eq30} \\
R &=& R^{(4)} -\frac{1}{4}\phi^{2} F_{ij}F^{ij} - \frac{2}{\phi}
\nabla^{2} \phi \label{eq31}
\end{eqnarray}
where $ R^{(4)}_{ij} $ and $ R^{(4)} $ are Ricci tensor and Ricci
scalar of $4D$ submanifold respectively with $F_{\mu\nu}$ being
the usual electromagnetic field tensor. Defining the $4D$
energy-momentum tensor and tension as \cite{james}
\begin{eqnarray}
\tau^{(4)}_{ij} =\frac{1}{\phi}  \tau_{ij} \nonumber \\
\label{eq32}\\ \sigma^{(4)} = \frac{1}{\phi}  \sigma \nonumber
\end{eqnarray}
and using the junction condition (\ref{eq227}) together with
equations (\ref{eq28}-\ref{eq31}) and equation (\ref{eq25}) result
in
\begin{eqnarray}
G^{(4)}_{ij} = K^{2}_{(4)} \tau^{(4)}_{ij} + \phi^2 \Sigma_{ij} +
\frac{1}{\phi^{2}}\Theta_{ij} - \frac{\epsilon}{4} K^{4}_{(B)}
\phi^{2} \Pi_{ij} \label{eq33} \\
6\nabla^{2} \phi -2 M^{2}_{H} \phi - \lambda \phi^{3} -
\frac{3}{2}\phi^3F_{ij}F^{ij} + 2 K^{2}_{(4)} \phi \left( 2{\cal
L}^{(4)}_{m} - \tau^{(4)} \right) \nonumber\\ + \frac{3}{8}
\epsilon K^{4}_{(B)} \phi^{3} \left( \tau^{(4)} - 3 {\cal
L}^{(4)}_{m} \right) {\cal
L}^{(4)}_{m} = 0 \label{eq34}\\
\nabla^{i} F_{ij} + \frac{3}{\phi}F_{ij}\nabla^{j}\phi = 0
\label{eq35}
\end{eqnarray}
where
\begin{eqnarray}
\Sigma_{ij} = \frac{1}{2} \left( F^{l}_{i} F_{lj} - \frac{1}{4}
F^{mn} F_{mn} g^{(4)}_{ij} \right) \label{eq36}
\end{eqnarray}
is the energy-momentum tensor of the Maxwell field and
\begin{eqnarray}
\Theta_{ij} = \phi \nabla_{i}\nabla_{j} \phi - \phi \nabla^{2}\phi
g^{(4)}_{ij} + \left( \frac{1}{2} M^{2}_{H} \phi^{2} +
\frac{\lambda}{4}\phi^{4} \right)g^{(4)}_{ij} \label{eq37}
\end{eqnarray}
is the corresponding equation for the Higgs field with the local
quadratic energy momentum correction being given by
\begin{eqnarray}
\Pi_{ij} &=& \frac{1}{16} \left[ {\cal
L}^{(4)}_{m}\tau^{(4)}g^{(4)}_{ij} - {\cal L}^{(4)}_{m}
\tau^{(4)}_{ij} - \frac{3}{2} \left({\cal L}^{(4)}_{m}\right)^2
g^{(4)}_{ij} - \tau^{(4)} \tau^{(4)}_{ij} + \frac{1}{2}
\left(\tau^{(4)}\right)^2 g^{(4)}_{ij}
+ 4 \tau^{(4)}_{im} \tau^{(4)m}_{j}\right. \nonumber \\
\nonumber \\ &-& \left. 2 \tau^{(4)mn} \tau^{(4)}_{mn}
g^{(4)}_{ij} \right] \label{eq38}
\end{eqnarray}
Also the $4D$ gravitational constant, the Higgs mass, the coupling
constant and the effective cosmological constant are respectively
given by
\begin{eqnarray}
K^{2}_{(4)} &=& \phi \left[ K^{2}_{(b)} - \frac{3 \epsilon}{8}
K^{4}_{(B)}\sigma \right] \label{eq39}\\
M^{2}_{H} &=& - 2 \left[ \frac{3}{4} \Lambda^{(B)} K^{2}_{(B)} +
\sigma
K^{2}_{(b)} \right]\label{eq40}\\
\lambda &=& \frac{3 \epsilon }{8}\ K^{4}_{(B)}
\left(\sigma^{(4)}\right)^2
\label{eq41}\\
\Lambda^{(4)} &=& \frac{1}{2}M^2_{H} - \frac{1}{4}\lambda \phi^2.
\label{eq42}
\end{eqnarray}
All these quantities have to be evaluated in the limit $ s = 0^{+}
$. From quantum field theory we know that the coupling constant of
the Higgs field must be positive, hence $ \epsilon = 1 $, {\it
i.e.} the large extra dimension must be spacelike. The present
value of the Higgs field is of order $ \phi \sim 10^{25}
\mbox{cm}^{-1}\sim \frac{1}{r}$, where $r$ is the radius of the
compact dimension. This means that $ r \sim 10^{-25} \mbox{cm} $
which is compatible with the cylinder condition. As can be seen
from equation (\ref{eq39}), the gravitational constant of the $4D$
spacetime $K_{(4)}$ is expressed in terms of the gravitational
constants of the brane and the bulk, the latter determining the
coupling constant $\lambda$ through equation (\ref{eq41}). Now, if
the bulk space did not exist, the coupling constant $\lambda$
would disappear and the $4D$ cosmological constant $\Lambda^{(4)}$
would not agree with the present observations as it is solely
expressed in terms of the Higgs mass, equation (\ref{eq42}).
However, in the presence of the bulk space, the $4D$ cosmological
constant is modified by the addition of the term $\lambda\phi^2$
which would modify the value of $\Lambda^{(4)}$ in such a way as
to make it more in line with the present observations. It is worth
noticing at this point that the absence of the large extra
dimension would render the theory ineffective in that
$\Lambda^{(4)}$ cannot be modified and would retain a large value
proportional to the Higgs mass, contrary to the present
observations.  As can be seen from equation (\ref{eq39}), the
existence of the small extra dimension is the cause of the
variations in our $4D$ gravitational constant.  We also note that
as equation (\ref{eq33}) shows, the local quadratic
energy-momentum corrections are now multiplied by a factor $
\phi^2 $, making the magnitude of these corrections dependent on
the scale of the small extra dimension.
\section{Localization of the gauge fields}
The localization of the gauge fields \cite{rubakov} implies that
gauge interactions should be confined to the brane. Assuming that
the gauge interactions are concomitant with the quantum
fluctuations of the geometry, it follows that equations
(\ref{eq7}) and (\ref{eq8}) must also be compatible with the
confinement of the gauge fields. In the Gauss-Codazzi equations,
the basic variables are $ g_{\alpha\beta} $ and $ K_{\alpha\beta}
$ and these quantities are influenced by the perturbations
according to equations (\ref{eq6})  and (\ref{eq9}) respectively.
From equation (\ref{eq6}) we have
\begin{eqnarray}
g_{i4} &=& A^{'}_{i}\phi^{'2} = \bar{g}_{i4} - 2s\bar{K}_{i4} +
s^2
\bar{g}^{\mu\nu} \bar{K}_{i\mu}\bar{K}_{4\nu} \nonumber \\
\label{eq43} \\ g_{44} &=& \phi^{'2} = \bar{g}_{44} -
2s\bar{K}_{44} + s^2 \bar{g}^{\mu\nu} \bar{K}_{4\mu}\bar{K}_{4\nu}
\nonumber
\end{eqnarray}
which represents the variations of various components of our $5D$
brane metric. Also from  the Junction conditions (\ref{eq227}) we
have
\begin{eqnarray}
\bar{K}_{i4} = A_{i}\bar{K}_{44} \label{eq44}.
\end{eqnarray}
Now from equations (\ref{eq43}) and (\ref{eq44}) one can see that
\begin{eqnarray}
A^{'}_{i} = A_{i} \label{eq45}
\end{eqnarray}
implying that $A_{i}$'s are localized in the sense that they
remain unchanged independently of the quantum fluctuations of the
brane-world. One should note that equation (\ref{eq45}) could not
be derived if the junction conditions were not assumed to hold,
that is, the junction conditions are closely related to the
localization of the gauge fields.

\section{Conclusions}
In this paper, we have studied the question of the unification of
the Higgs and Maxwell fields in a model where a brane with a
compact dimension is embedded in a bulk space having a constant
curvature. Such a setup with its extra small and large dimensions
enabled us to construct the Higgs and gauge fields  purely in
terms of the geometric quantities. We have also noted that within
the context of the present model, there are three scales of
energy;  the unification of the fundamental forces in the bulk,
the unification of the gauge fields on the brane arising from
having an extra compact dimension and the energy scale of gravity
in four dimensions. To account for the gauge fields other than
Maxwell's, one would, in principle, increase the number of the
compact dimensions of the brane in a straightforward manner.
\vspace{5mm}\noindent\\
{\bf Acknowledgements}\vspace{2mm}\\ The authors would like to
thank A. Ahmadi and P. Moyassari for useful discussions. SJ would
also like to thank the research council of Shahid Beheshti
university for financial support.


\begin{thebibliography}{99}

\bibitem{arkani} N. Arkani-Hamed, S. Dimopoulos and G. Dvali,
Phys. Lett. B {\bf 429}, 263  (1998);\\
I. Antoniadis, N. Arkani-Hamed, S. Dimopoulos and G. Dvali, Phys.
Lett. B {\bf 436}, 257 (1998).
\bibitem{randal}L. Randall and R. Sundrum,
Phys. Rev. Lett. {\bf 83}, 4690 (1999);\\L. Randall and R.
Sundrum, Phys. Rev. Lett. {\bf 83}, 3370 (1999).
\bibitem{burdman} G. Burdman and Y. Nomura, Nucl. Phys. B
{\bf 656}, 3 (2003).
\bibitem{benakli} K. Benakli and M. Quiros, C. R. Physique {\bf 4}, 363 (2003).
\bibitem{eis} L. P. Eisenhart, {\it Riemannian Geometry, Princeton U.
P.}, Princeton, N. J. (1966);\\
Y. Aminov, {\it The Geometry of Submanifolds, Gordon and Breach,
Sience Publishers} (2001).
\bibitem{maia} M. D. Maia and Edmundo M. Monte, Phys. Lett. A {\bf 297}, 9 (2002).
\bibitem{finn} I. K. Vehus and F. Ravndal, hep-ph/0210292.
\bibitem{james} J. M. Cline and G. Vinet, Phys. Rev. D {\bf 68}, 025015
(2003).
\bibitem{rubakov} V. A. Rubakov, Phys. Usp {\bf 44}, 871 (2001).
\end{thebibliography}
\end{document}